\UseRawInputEncoding

\documentclass[pra,aps,twocolumn,epsfig,superscriptaddress,showpacs]{revtex4}
\usepackage{amsmath}
\usepackage{amssymb}
\usepackage[dvips]{graphicx}
\usepackage{dcolumn}
\usepackage{bm}
\usepackage[colorlinks=false,dvipdfm]{hyperref} 
\usepackage{setspace}
\usepackage{amsfonts}
\usepackage{rotating}
\usepackage{makeidx}
\usepackage{amsthm}
\usepackage{bbold}
\usepackage{float}

\begin{document}

\title{Probabilistic resumable quantum teleportation in high dimensions \footnote{Chin. Phys. B 31, 030302 (2022)}}

\author{Xiang Chen}
\affiliation{Department of Physics, School of Science, Tianjin University, Tianjin 300072, China}

\author{Jin-Hua Zhang}
\affiliation{School of Mathematical Sciences, Capital Normal University, Beijing 100048, China}
\affiliation{Department of Physics, Xinzhou Teacher's University, Xinzhou 034000, China}

\author{Fu-Lin Zhang}
\email[Corresponding author: ]{flzhang@tju.edu.cn}
\affiliation{Department of Physics, School of Science, Tianjin University, Tianjin 300072, China}

\begin{abstract}
Teleportation is a quantum information processes without classical counterparts, in which the sender can disembodied transfer unknown quantum states to the receiver.
In probabilistic teleportation through a partial entangled quantum channel, the transmission is exact (with fidelity 1),
but may fail in a probability and  the initial state is destroyed simultaneously.
We propose a scheme for nondestructive probabilistic teleportation of high-dimensional quantum states.
With the aid of an ancilla in the hands of the sender, the initial quantum information can be recovered when teleportation fails.
The ancilla acts as a quantum apparatus to measure the sender's subsystem.
Erasing the information recorded in it can resume the initial state.
\end{abstract}

%
 \maketitle

\section{Introduction}\label{Intro}

Quantum teleportation is one of the processes unique to the quantum information, which serves as an important example for the most intriguing uses of entanglement \cite{Book,bennett1993teleporting,brunner2005entanglement,RevModPhys.81.865}.
It has been widely studied both theoretically and experimentally since it was put forwards  \cite{1999Probabilistic,Banaszek2000Optimal,Roa2003Optimal,2003Optimal,PRA2015RoaTele,Ctele,PhysRevA.90.052305,zhang2012assisted,ZhangEPL2017,Zhang2021,HUANG2020Quantum,2019Quantum,2020Experimental,1997Experimental,Boschi1997Experimental,Furusawa2019Unconditional,Gottesman1999Demonstrating,2013Efficient,2014Unconditional,2014Quantum}.
%
One reason it is gaining so much attraction is that,
the original scheme and its various extensions 
play key roles in several contexts in quantum communication, including quantum repeaters, quantum networks and cryptographic conferences \cite{repeaters,PhysRevA.54.2651,PhysRevA.57.822,networks,QNet}.

In the original and also the simplest version of teleportation \cite{bennett1993teleporting},  Alice (the sender) can transfer an unknown state from a qubit to Bob (the receiver), with fidelity 1 and successful probability 1, without physical transmission of the qubit itself.
The key ingredient in the protocol is a two-qubit Bell state shared by them as a quantum channel.
Alice makes a joint measurement on the two qubits in her hands, projecting them onto one of the four Bell states with equal probability.
Through a classical channel, Alice sends Bob two classical bits to inform {him of her outcome}.
According to the classical information, Bob can perform appropriate unitary operations on his qubit to {rebuild} Alice's initial state, and thereby {accomplishes} the teleportation.

A variant of the process called probabilistic teleportation was proposed
based upon the consideration of that a quantum channel may be prepared in a partially entangled pure state in practice
\cite{1999Probabilistic,Banaszek2000Optimal,Roa2003Optimal,2003Optimal,PRA2015RoaTele}.
Such teleportation is exact (with fidelity 1), but may sometimes fail as the the price to pay for the fidelity.
In principle, Alice's joint measurement is destructive, and it is generally assumed that the information {encoded} in the state to be teleported is lost when the teleportation fails.
In their recent work \cite{PRA2015RoaTele}, Roa and Groiseau presented a nondestructive scheme for probabilistic teleportation by introducing an ancillary qubit.
This avoids losing the initial information and offers the chance to repeat the teleportion process.
The nondestructive scheme has been extended in many branches, including bidirectional teleportation \cite{QINP2017}, teleportation of an entangled state \cite{Entropy2019} and multihop scenario \cite{JOSAB2020a,JOSAB2020b}.

In this work, we present a general protocol for nondestructive probabilistic teleportation of high-dimensional quantum states,
in which Alice can resume her initial state when the teleportation fails.
The motivation for this work is not only to extend the study of Roa and Groiseau to high-dimensional systems, but also based on the following considerations.
In theory, high-dimensional channels, especially the partially entangled ones, have more rich entanglement properties.
Our protocol provides a sample to study the cooperative relationship among the entanglement, joint measurement and classical information in a quantum information process.
In practice, {teleporting} high-dimensional states is required in the task to completely rebuild the quantum states of a a real particle remotely.
And, recent experiment progresses \cite{2019Quantum,2020Experimental} show the possibility of {implementing} our present protocol in optical systems.
We make a remark here that, although Fu \textit{et al.}  \cite{JOSAB2020a,JOSAB2020b} give a version of nondestructive probabilistic teleportation in high dimension using an auxiliary particle in Bob's hand, we maintain the approach in \cite{PRA2015RoaTele} with an ancilla belonging to Alice.
In this case,   only the sender, Alice, is required {to have} the ability of bipartite operations, such as generation and measurement of entangled states.

In {addition}, one can notice the similarity of our protocol and the quantum two-path interference experiments \cite{PRL2008twopath}.
In the latter, interference fringes vanish when the which-path information is  acquired by a detector,
 while reappear when the detector returnes the information to the particle.
The ancilla in our protocol  can be regarded as a  quantum apparatus to measure Alice's system.
The initial state is recovered by erasing the information it records after an unambiguous quantum state discrimination.

%

 \section{Teleportation of  a qutrit}	\label{qutrit}

Let us start with  the teleportation of  a qutrit (a three-level quantum system) .
Suppose Alice wishes to teleport to Bob an arbitrary qutrit state as
\begin{equation}\label{qutritST}
\left\vert \phi\right\rangle _{1}= \alpha_{0}\left\vert0\right\rangle_{1}+\alpha_{1}\left\vert 1\right\rangle_{1} +\alpha_{2}\left\vert2\right\rangle _{1} ,
\end{equation}
with $\left\vert \alpha_{0}\right\vert ^{2}+\left\vert\alpha_{1}\right\vert^{2}+\left\vert \alpha_{2}\right\vert ^{2}=1$.
The participants share a two-qutrit entangled state as the quantum channel,
\begin{equation}\label{channel3}
\left\vert \Phi\right\rangle _{23}=b_{0}\left\vert 00\right\rangle_{23}+b_{1}\left\vert 11\right\rangle _{23}+b_{2}\left\vert 22\right\rangle
_{23},
\end{equation}
with $\left\vert b_{0}\right\vert ^{2}+\left\vert b_{1}\right\vert ^{2}+\left\vert b_{2}\right\vert ^{2}=1$.
Without loss of generality, we assume {that} the Schmidt coefficients are non-negative real numbers and $  b_{0}  \leq  b_{1} \leq  b_{2}  $.
Here, the three qutrits are identified by the subscripts $1,2$ and $3$ respectively, and
qutrits $1$ and $2$ belong to Alice while qutrit $3$  is in the hands of Bob.
The total state of the three-qutrit system can be written as,
\begin{align}\label{qutritTT}
\left\vert \Psi\right\rangle _{123} & =\left\vert\phi\right\rangle_{1}\otimes\left\vert \Phi\right\rangle _{23}  \notag \\
& =\frac{1}{3}\bigr[\   \left\vert \psi_{00}\right\rangle_{12}(\alpha_{0}\left\vert0\right\rangle +\alpha_{1}\left\vert1\right\rangle +\alpha_{2}\left\vert2\right\rangle )_{3}  \notag \\
&\ \ \ \ \ \   +\!\left\vert \psi_{10}\right\rangle _{12}(\alpha_{0}\left\vert0\right\rangle +\alpha_{1}\omega^{2}\left\vert 1\right\rangle+\alpha_{2}\omega\left\vert 2\right\rangle )_{3}  \notag \\
&\ \ \ \ \ \   +\!\left\vert \psi_{20}\right\rangle_{12}(\alpha_{0}\left\vert0\right\rangle +\alpha_{1}\omega\left\vert1\right\rangle +\alpha_{2} \omega^{2}\left\vert 2\right\rangle )_{3}  \notag\\
&\ \ \ \ \ \   +\!\left\vert \psi_{01}\right\rangle _{12}(\alpha_{1}\left\vert0\right\rangle +\alpha_{2}\left\vert 1\right\rangle
+\alpha_{0}\left\vert2\right\rangle )_{3} \notag  \\
&\ \ \ \ \ \  +\!\left\vert \psi_{11}\right\rangle_{12}(\alpha_{1}\left\vert0\right\rangle +\alpha_{2}\omega^{2}\left\vert1\right\rangle +\alpha _{0}\omega\left\vert 2\right\rangle )_{3}   \\
&\ \ \ \ \ \   +\!\left\vert \psi_{21}\right\rangle _{12}(\alpha_{1}\left\vert0\right\rangle +\alpha_{2}\omega\left\vert 1\right\rangle+\alpha_{0}\omega^{2}\left\vert 2\right\rangle )_{3}  \notag \\
&\ \ \ \ \ \  +\!\left\vert \psi_{02}\right\rangle_{12}(\alpha_{2}\left\vert0\right\rangle +\alpha_{0}\left\vert1\right\rangle +\alpha_{1}\left\vert 2\right\rangle )_{3}  \notag \\
&\ \ \ \ \ \   +\!\left\vert \psi_{12}\right\rangle _{12}(\alpha_{2}\left\vert0\right\rangle +\alpha_{0}\omega^{2}\left\vert 1\right\rangle+\alpha_{1}\omega\left\vert 2\right\rangle )_{3}  \notag \\
&\ \ \ \ \ \  +\!\left\vert \psi_{22}\right\rangle_{12}(\alpha_{2}\left\vert0\right\rangle +\alpha_{0}\omega\left\vert1\right\rangle +\alpha_{1} \omega^{2}\left\vert 2\right\rangle )_{3}  \  \bigr], \notag \ \ \ \ \ \
\end{align}
where  $\omega=e^{i \frac{ 2\pi}{3}}$ is the triple root,
and $\left\vert \psi_{nm}\right\rangle_{12}$  are nine linearly independent two-qutrit states,  equivalent to the channel (\ref{channel3}) under local unitary transformations, as follows,
\begin{align}\label{qutritbasis}
\left\vert \psi_{00}\right\rangle _{12} & =(b_{0}\left\vert00\right\rangle+b_{1}\left\vert 11\right\rangle +b_{2}\left\vert
22\right\rangle )_{12},  \notag \\
\left\vert \psi_{10}\right\rangle _{12} & =(b_{0}\left\vert00\right\rangle+b_{1}\omega\left\vert 11\right\rangle
+b_{2}\omega^{2}\left\vert 22\right\rangle )_{12},  \notag \\
\left\vert \psi_{20}\right\rangle _{12} & =(b_{0}\left\vert00\right\rangle+b_{1}\omega^{2}\left\vert 11\right\rangle
+b_{2}\omega\left\vert 22\right\rangle )_{12},  \notag \\
\left\vert \psi_{01}\right\rangle _{12} & =(b_{0}\left\vert10\right\rangle+b_{1}\left\vert 21\right\rangle +b_{2}\left\vert
02\right\rangle )_{12},  \notag \\
\left\vert \psi_{11}\right\rangle _{12} & =(b_{0}\left\vert10\right\rangle+b_{1}\omega\left\vert 21\right\rangle
+b_{2}\omega^{2}\left\vert 02\right\rangle )_{12},   \\
\left\vert \psi_{21}\right\rangle _{12} & =(b_{0}\left\vert10\right\rangle+b_{1}\omega^{2}\left\vert 21\right\rangle
+b_{2}\omega\left\vert 02\right\rangle )_{12},  \notag \\
\left\vert \psi_{02}\right\rangle _{12} & =(b_{0}\left\vert20\right\rangle+b_{1}\left\vert 01\right\rangle +b_{2}\left\vert
12\right\rangle )_{12},  \notag \\
\left\vert \psi_{12}\right\rangle _{12} & =(b_{0}\left\vert20\right\rangle+b_{1}\omega\left\vert 01\right\rangle
+b_{2}\omega^{2}\left\vert 12\right\rangle )_{12},  \notag \\
\left\vert \psi_{22}\right\rangle _{12} & =(b_{0}\left\vert20\right\rangle+b_{1}\omega^{2}\left\vert 01\right\rangle
+b_{2}\omega\left\vert 12\right\rangle )_{12}. \notag
\end{align}
The vector of qutrit $3$ in each term (each line) of the total state (\ref{qutritTT}) is equivalent to the initial state (\ref{qutritST}) under a unitary {transformation}.
However,  the corresponding  states of qutrits $1$ and $2$ $\left\vert \psi_{nm}\right\rangle_{12}$  are not orthogonal to each other.
To teleportate the initial state exactly, an option is that, Alice performs an unambiguous quantum state discrimination process \cite{Banaszek2000Optimal,Roa2003Optimal,zhang2012assisted,ZhangEPL2017},
to distinguish the nine states with no error but 
a probability of failures.
As in the standard teleportation through a maximal entangled quantum channel  \cite{2019Quantum,2020Experimental}, Bob can rebuild the initial state on his qutrit by performing appropriate unitary operations according to Alice's outcome.
In another scheme, Alice still performs a joint measurement in nine maximally entangled basis,
while Bob needs an extracting quantum state process \cite{1999Probabilistic}.
Here, we follow the former scheme,  which only requires Alice's ability to manipulate two or more particles.

 \begin{figure}
 \includegraphics[width=7.5cm]{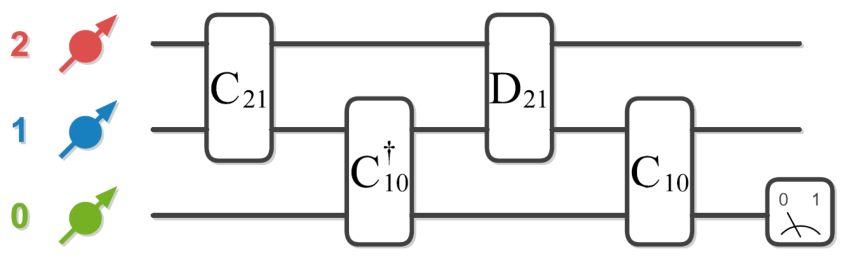}
\caption{
(Color online) Alice's sequential operations on her subsystems:
(1) The GCNOT gate $C_{21}$  factorizes the states of $1$ and $2$  to be discriminated;
(2) The apparatus  $0$ measures $1$ by using  $C_{10}^\dag$;
(3) Alice performs a joint transformation $D_{21}$ on  $1$ and $2$ to discriminate the states of $2$;
(4) The GCNOT gate $C_{10}$  erases the information of  $1$ recorded in $0$;
(5) Alice measures $0$ to divide the procedure into success and failure parts.
}\label{Fig1}
\end{figure}

For the sake of brevity, before putting forward our final total state in this section, we only show the states in the hands of Alice.
We show the  sequential operations of Alice in Fig. \ref{Fig1}.
To discriminate the nine two-qutrit states ({\ref{qutritbasis}}), Alice factorizes the states by applying  a generalized controlled-NOT (GCNOT) gate onto her bipartite system and obtain
\begin{align}\label{qutritphinm}
\mathcal{C}_{21}\left\vert \psi_{nm}\right\rangle_{12}= \left\vert m \right\rangle_{1} ( b_{0}\left\vert 0\right\rangle +b_{1}\omega^{n}\left\vert 1\right\rangle +b_{2}\omega^{2n}\left\vert
2\right\rangle)_2,
\end{align}
with $m,n=0,1,2$.
Here, we define the GCNOT gate $C_{ij}$ acting on qutrits $i$ and $j$ as
\begin{align}
\mathcal{C}_{ij}  &  =\left\vert 0\right\rangle _{i}\left\langle 0\right\vert  \otimes \openone_j + \left\vert 1\right\rangle _{i}\left\langle 1\right\vert \otimes \mathcal{V}_j+\left\vert 2\right\rangle _{i}\left\langle 2\right\vert \otimes
\mathcal{V}_j^{\dag},
\end{align}
where $\openone_j$ is the identity of $j$ and
$\mathcal{V}_j =\left\vert 0\right\rangle_j \left\langle 1\right\vert +\left\vert1\right\rangle_j \left\langle 2\right\vert +\left\vert 2\right\rangle_j \left\langle 0\right\vert$.
It shifts the target $j$  clockwise or anticlockwise when the control qutrit $i$ is a $\left\vert 1\right\rangle$ or $\left\vert 2\right\rangle$.

To distinct the nine states in (\ref{qutritphinm}), Alice can perform a von Neumann measurement on qutrit $1$ followed by an unambiguous quantum state discrimination on qutrit $2$.
However, these operations destroy the initial state of qutrit $1$ even though the discrimination fails.
Here, following the protocol of Roa and Groiseau \cite{PRA2015RoaTele}, we introduce an extra auxiliary qutrit $0$ which acts as a quantum apparatus to measure qutrit $1$.
The key point is that, when we erase the information of qutrit $1$ recorded in the ancilla, the initial state of  qutrit $1$  recoveries when discrimination fails.
Alice applies the  inverse of the GCNOT gate on the ancilla initial in $\left\vert 0\right\rangle_0$ and qutrit $1$, and obtain the nine states in her hands as
\begin{align}
\mathcal{C}_{10}^{^{\dag}} \left\vert 0\right\rangle_0 \left\vert m \right\rangle_{1} \left\vert \tau_{n}\right\rangle_2=  \left\vert m \right\rangle_0 \left\vert m \right\rangle_{1} \left\vert \tau_{n}\right\rangle_2,
\end{align}
where $\left\vert \tau_{n}\right\rangle_2 = b_{0}\left\vert 0\right\rangle_2 +b_{1}\omega^{n}\left\vert 1\right\rangle_2 +b_{2}\omega^{2n}\left\vert2\right\rangle_2$ and $m,n=0,1,2$.
Then, qutrit $1$ {serves} as an ancilla in the unambiguous quantum state discrimination of  qutrit $2$.
Let  us define two unitary transformations on qutrit $2$ to be
\begin{align}
\begin{split}
\mathcal{U}_1 & =\frac{b_{0}}{b_{1}} \openone_1 +\sqrt{1-\left(\frac{b_{0}}{b_{1}}\right)^{2}} \mathcal{V}_1,   \\
\mathcal{W}_1 &=\frac{b_{0}}{b_{2}} \openone_1 +\sqrt{1-\left(\frac{b_{0}}{b_{2}}\right)^{2}} \mathcal{V}_1,
\end{split}
\end{align}
and a controlled-unitary operation
\begin{align}
\mathcal{D}_{21}   =\left\vert 0\right\rangle _{2}\left\langle 0\right\vert\otimes \openone_{1}+\left\vert 1\right\rangle _{2}\left\langle 1\right\vert\otimes  \mathcal{U}_1 +\left\vert 2\right\rangle _{2}\left\langle 2\right\vert\otimes \mathcal{W}_1.
\end{align}
Alice performs it on qutrits $1$ and $2$ and obtains
\begin{align}
\begin{split}
\mathcal{D}_{21} & \left\vert m \right\rangle_0  \left\vert m \right\rangle_{1} \left\vert \tau_{n}\right\rangle_2 =  \left\vert m \right\rangle_0 \biggr[ \sqrt{3}b_0 \left\vert m \right\rangle_{1} \left\vert \varkappa_{n}\right\rangle _{2} \\&
   +  \left\vert m\oplus_{3}2 \right\rangle_{1}( \sqrt{b_1^2 - b_0^2}\omega^{n} \left\vert 1 \right\rangle_{2} + \sqrt{b_2^2 - b_0^2}\omega
^{2n} \left\vert 2 \right\rangle_{2} ) \biggr],  \ \ \ \ \ \ \
\end{split}
\end{align}
where $\oplus_{3}$ denotes modulo $3$ addition, and $\left\vert \varkappa_{n}\right\rangle  =\frac{1}{\sqrt{3}}(\left\vert0\right\rangle _{2}+\omega^{n}\left\vert 1\right\rangle _{2}+\omega^{2n}\left\vert 2\right\rangle _{2})$.
In the above form, the nine states
 $ \left\vert m \right\rangle_0  \left\vert m \right\rangle_{1} \left\vert \varkappa_{n}\right\rangle _{2} $
 corresponding to successful discrimination are orthogonal to each other, and  orthogonal to the states with qutrits $0$ and $1$ in
  $ \left\vert m \right\rangle_0 \left\vert m\oplus_{3}2 \right\rangle_{1} $.
The latter nine are for the failure of  discrimination as they are not linearly independent.

The final step of Alice's unitary operations is to erase the information measured by qutrit $0$.
 Applying the GCNOT gate $\mathcal{C}_{10}$, Alice can obtain
\begin{align}
\begin{split}
& \mathcal{C}_{10} \left\vert m \right\rangle_0  \left\vert m \right\rangle_{1} = \left\vert 0 \right\rangle_0  \left\vert m \right\rangle_{1}, \\
& \mathcal{C}_{10}\left\vert m \right\rangle_0 \left\vert m\oplus_{3}2 \right\rangle_{1}= \left\vert 1 \right\rangle_0 \left\vert m \oplus_{3}2 \right\rangle_{1}.
\end{split}
 \end{align}
The information recorded on the apparatus  qutrit $0$ can be erased partially, as it returns $\left\vert 0 \right\rangle_0$ in the terms corresponding to successful discrimination but becomes $\left\vert 1 \right\rangle_0$ for the case of failure.
This divides the total four-qutrit state into two parts as
\begin{align}
\begin{split}
\left\vert \Delta\right\rangle
&  =\frac{b_{0}}{\sqrt{3}}\left\vert 0\right\rangle _{0} \biggr[
\left\vert0\right\rangle _{1} \left(\left\vert \! \varkappa_{0} \!  \right\rangle \left\vert \!  \phi_{00}\!  \right\rangle   \!  + \! \left\vert  \! \varkappa_{1} \! \right\rangle  \left\vert \! \phi_{10}\!   \right\rangle \!  +\!  \left\vert\!  \varkappa_{2}\! \right\rangle  \left\vert \! \phi_{20}\! \right\rangle\right)_{23}\\
&\ \ \ \ \ \ \ \  \ \ \ \ \   +\left\vert1\right\rangle _{1} \left(\left\vert \!  \varkappa_{0}\!  \right\rangle \left\vert\!  \phi_{01}\! \right\rangle  \!   +\! \left\vert \! \varkappa_{1}\! \right\rangle  \left\vert\!  \phi_{11} \!  \right\rangle \! +\!  \left\vert \! \varkappa_{2}\! \right\rangle  \left\vert\! \phi_{21}\! \right\rangle\right)_{23}\\
&\ \ \ \ \ \ \ \  \ \ \ \ \    +\left\vert2\right\rangle _{1} \left(\left\vert \!  \varkappa_{0}\!  \right\rangle \left\vert \!  \phi_{02}\!  \right\rangle\!     + \! \left\vert\!  \varkappa_{1}\!  \right\rangle  \left\vert \! \phi_{12}\!  \right\rangle \! +\! \left\vert \! \varkappa_{2}\! \right\rangle  \left\vert\!  \phi_{22}\! \right\rangle\right)_{23}
 \biggr] \\
& +\left\vert 1\right\rangle _{0}\biggr[
\sqrt{b_{1}^{2}\! -\! b_{0}^{2}}\left\vert\!  \phi_{02}\! \right\rangle \left\vert 1\right\rangle  \left\vert1\right\rangle
  \! +\! \sqrt{b_{2}^{2}\! -\! b_{0}^{2}}\left\vert\! \phi_{00}\! \right\rangle \left\vert2\right\rangle \left\vert 2\right\rangle
   \biggr]_{123}. \ \ \
\end{split}
\end{align}
Here $\left\vert  \phi_{nm} \right\rangle $ are the states of qutrit $3$ multiplied to $\left\vert  \psi_{nm} \right\rangle_{12} $ in (\ref{qutritTT}) as
\begin{align}
\begin{split}
\left\vert \phi_{00}\right\rangle    & =(\alpha_{0}\left\vert0\right\rangle +\alpha_{1}\left\vert 1\right\rangle +\alpha_{2}\left\vert
2\right\rangle ) ,\\
\left\vert \phi_{10}\right\rangle    & =(\alpha_{0}\left\vert0\right\rangle +\alpha_{1}\omega^{2}\left\vert 1\right\rangle +\alpha
_{2}\omega\left\vert 2\right\rangle ), \\
\left\vert \phi_{20}\right\rangle    & =(\alpha_{0}\left\vert0\right\rangle +\alpha_{1}\omega\left\vert 1\right\rangle +\alpha_{2}
\omega^{2}\left\vert 2\right\rangle ) ,\\
\left\vert \phi_{01}\right\rangle    & =(\alpha_{1}\left\vert0\right\rangle +\alpha_{2}\left\vert 1\right\rangle +\alpha_{0}\left\vert
2\right\rangle ) ,\\
\left\vert \phi_{11}\right\rangle    & =(\alpha_{1}\left\vert0\right\rangle +\alpha_{2}\omega^{2}\left\vert 1\right\rangle +\alpha
_{0}\omega\left\vert 2\right\rangle ) ,\\
\left\vert \phi_{21}\right\rangle    & =(\alpha_{1}\left\vert0\right\rangle +\alpha_{2}\omega\left\vert 1\right\rangle +\alpha_{0}
\omega^{2}\left\vert 2\right\rangle ) ,\\
\left\vert \phi_{02}\right\rangle    & =(\alpha_{2}\left\vert0\right\rangle +\alpha_{0}\left\vert 1\right\rangle +\alpha_{1}\left\vert
2\right\rangle ), \\
\left\vert \phi_{12}\right\rangle    & =(\alpha_{2}\left\vert0\right\rangle +\alpha_{0}\omega^{2}\left\vert 1\right\rangle +\alpha
_{1}\omega\left\vert 2\right\rangle ), \\
\left\vert \phi_{22}\right\rangle    & =(\alpha_{2}\left\vert0\right\rangle +\alpha_{0}\omega\left\vert 1\right\rangle +\alpha_{1}
\omega^{2}\left\vert 2\right\rangle ).
\end{split}
\end{align}
Then, the state to be teleported is encoded in qutrit $3$ in  the first part of $\left\vert \Delta\right\rangle$, but in qutrit $1$ in the second part.

Alice performs a  von Neumann measurements on qutrit $0$ in the standard basis.
It projects qutrit $0$  to $\left\vert 0\right\rangle_{0}$ in a probability $3\left\vert b_{0}\right\vert ^{2}$.
Then, she measures the nine orthogonal direct product states $\left\vert 0 \right\rangle_0  \left\vert m \right\rangle_{1}$,
and informs Bob to perform  appropriate unitary operations on his qutrit $3$ to rebuild  the initial state, and thereby the teleportation succeeds.
On the contrary, the qutrit $0$ could be  projected to $\left\vert 1\right\rangle _{0}$ in a probability $1-3\left\vert b_{0}\right\vert ^{2}$,
which means the failure of the teleportation.
Alice can recover the  state $\left\vert \phi\right\rangle _{1}$ in qutrit $1$, by performing a joint  unitary transformation  $\mathcal{B}_{21}    = \openone_1 \otimes (\left\vert 0\right\rangle _{2}\left\langle 0\right\vert +\left\vert 2\right\rangle _{2}\left\langle 2\right\vert   ) + \mathcal{V}_1  \otimes  \left\vert 1\right\rangle _{2}\left\langle 1\right\vert$ on qutrits $1$ and $2$.
Alternatively, Alice could perform a  von Neumann measurements on qutrit $2$ in the standard basis.
She obtains the initial state $\left\vert \phi\right\rangle _{1}$, by using the operation $\mathcal{V}_1$ on qutrit  $1$ when qutrit $2$ was projected to  $\left\vert 2\right\rangle _{2}$, while $\openone_1$ corresponding to $\left\vert 0\right\rangle _{2}$ and $\left\vert 1\right\rangle _{2}$.




%
  \section{General protocol}	\label{qudit}
Now we turn to the general protocol for teleporting a quNit (a $N$-level quantum system) through a partially entangled two-quNit quantum channel.
Since it is direct to extend the above scheme to
$N$-level systems, we present the following results in  compact general formulae in this part.
Let an initial quNit state to be
\begin{equation}\label{quNitST}
\left\vert \phi\right\rangle _{1}=
{\textstyle\sum\limits_{i=0}^{N-1}}
\alpha_{i}\left\vert i\right\rangle _{1},
\end{equation}
where $\alpha_{i=0,1,\cdots,N-1}$ are complex numbers satisfying the normalization condition
$%
{\textstyle\sum\nolimits_{i=0}^{N-1}}
\left\vert \alpha_{i}\right\vert^2 =1$. 
 Alice and Bob share a two-quNit entangled state as the quantum channel, which  can be generally written in the form of the Schmidt disposition as
\begin{equation}
\left\vert \Phi\right\rangle _{23}=
{\textstyle\sum\limits_{j=0}^{N-1}}
b_{j}\left\vert j\right\rangle _{2}\left\vert j\right\rangle _{3},
\end{equation}
with the real coefficient $0 \leq  b_{0}  \leq  b_{1}  \leq\ldots\leq  b_{N-1}  $ and
$
{ \sum_{j=0}^{N-1}}
  b_{j} ^{2}=1$.
The total state of the tripartite system is given by
\begin{align}\label{total state}
\left\vert \Psi\right\rangle _{123} &  =\left\vert \phi\right\rangle
_{1} \left\vert \Phi\right\rangle _{23}\nonumber\\
&  =\frac{1}{N}%
{ \sum_{m=0,n=0}^{N-1,N-1}}
 \left\vert \psi_{nm}\right\rangle _{12}     {\textstyle\sum\limits_{f=0}^{N-1}}\alpha_{f\oplus  m}e^{\frac{-i2\pi fn}{N}}\left\vert  f  \right\rangle _{3} ,  \ \
\end{align}
where  $\oplus $ denotes modulo $N$ addition,
and
$\left\vert \psi_{nm}\right\rangle _{12}$ are $N^2$  linearly independent bipartite states
\begin{equation}
\left\vert \psi_{nm}\right\rangle _{12}=
{\textstyle\sum\limits_{k=0}^{N-1}}
b_{ k }e^{\frac{i2\pi kn}{N}}\left\vert k\oplus  m\right\rangle_{1}\left\vert  k\right\rangle _{2}.
\end{equation}
In the above form {in Eq}. (\ref{total state}), the states  multiplied to $\left\vert  \psi_{nm} \right\rangle_{12} $  are equivalent to the  state (\ref{quNitST}) on quNit $3$ under unitary transformations as
\begin{align}
{\textstyle\sum\limits_{f=0}^{N-1}}\alpha_{f\oplus  m}e^{\frac{-i2\pi fn}{N}}\left\vert  f  \right\rangle _{3} = U_{3}^{(n,m)}\left\vert \phi\right\rangle _{3}
\end{align}
with
\begin{align}
U_{j}^{(n,m)} =
{\textstyle\sum\limits_{f=0}^{N-1}}
e^{\frac{-i2\pi fn}{N}}\left\vert f\right\rangle_j \left\langle
  f\oplus  m \right\vert,
\end{align}
and the subscript $j$ denoting the $j$th subsystem.

Alice's first two operations are to disentangle subsystems $1$ and $2$ in the states $\left\vert  \psi_{nm} \right\rangle_{12} $,
and to measure $2$ by using an auxiliary apparatus, quNit $0$ {initial} in $\left\vert 0\right\rangle _{0}$.
Here, we define the $N$-level  GCNOT gate as
 \begin{align}
C_{ij} =
{\textstyle\sum\limits_{y=0}^{N-1}}
\left\vert y\right\rangle _{i}\left\langle y\right\vert \otimes U_{j}^{(0,y)}.
\end{align}
The four-partite state becomes
\begin{align}
\left\vert \Omega\right\rangle  &  =C_{10}^{\dag}\biggr(\left\vert 0\right\rangle _{0}  C_{21}\left\vert \Psi\right\rangle_{123} \biggr)
\nonumber\\
&  =\frac{1}{N}%
 \! {\sum_{m=0,n=0}^{N-1,N-1}}  \!
 \left\vert mm\right\rangle _{01}
\left( \!
{\textstyle\sum\limits_{j=0}^{N-1}}
b_{j}e^{\frac{i2\pi jn}{N}}\left\vert j\right\rangle _{2}
 \!
 \right)
 \! U_{3}^{(n,m)}  \!  \! \left\vert \phi\right\rangle _{3} . \ \nonumber
\end{align}

To unambiguously discriminate the states of particle $2$, Alice applies a joint unitary transformation on $1$ and $2$
\begin{align}
D_{21}   &=
{\textstyle\sum\limits_{y=0}^{N-1}}
\left\vert y\right\rangle _{2}\left\langle y\right\vert \otimes \left(\frac{b_{0}}{b_{y}} \openone_1 +\sqrt{1-\left(\frac{b_{0}}{b_{y}}\right)^{2}} U_1^{(0,1)}  \right),\nonumber
\end{align}
obtains the whole system in the state
\begin{align}
\left\vert \Gamma\right\rangle  & =D_{21}\left\vert \Omega
\right\rangle \nonumber\\
 & =\frac{b_{0}}{N}%
 \! {\sum_{m=0,n=0}^{N-1,N-1}}  \!
 \left\vert mm\right\rangle _{01}
\left( \!
{\textstyle\sum\limits_{j=0}^{N-1}}
e^{\frac{i2\pi jn}{N}}\left\vert j\right\rangle _{2}
 \!
 \right)
 \! U_{3}^{(n,m)}  \!  \! \left\vert \phi\right\rangle _{3}    \nonumber\\
& +%
\! {\sum_{m=0,j=1}^{N-1,N-1}} \!
 \sqrt{\! b_{j}^{2}-b_{0}^{2} \! }\left\vert m\right\rangle _{0}
 \alpha_{ j \oplus m   }\left\vert m \oplus N\!-\!1  \right\rangle _{1} \left\vert jj\right\rangle _{23}. \nonumber
\end{align}
When Alice erases the  information in the auxiliary apparatus, $0$,
the changes of the quNit $1$ are  recorded as  $\left\vert1\right\rangle _{0}$, and the total state becomes
\begin{align}
\left\vert \Delta\right\rangle  & =C_{10}\left\vert \Gamma
\right\rangle \nonumber\\
&  = \left\vert  0\right\rangle _{0} \frac{b_{0}}{N}%
 \! {\sum_{m=0,n=0}^{N-1,N-1}}  \!
 \left\vert  m\right\rangle _{1}
\left( \!
{\textstyle\sum\limits_{j=0}^{N-1}}
e^{\frac{i2\pi jn}{N}}\left\vert j\right\rangle _{2}
 \!
 \right)
 \! U_{3}^{(n,m)}  \!  \! \left\vert \phi\right\rangle _{3}    \nonumber\\
&\ \ \  +%
\left\vert 1 \right\rangle _{0} \! {\sum_{ j=1}^{ N-1}} \!
 \sqrt{\! b_{j}^{2}-b_{0}^{2}  }
U_{1}^{(0,j\oplus1)}\left\vert \phi \right\rangle _{1} \left\vert jj\right\rangle _{23}. \nonumber
\end{align}
Obviously, it is divided into two parts of success and failure, which can be collapsed by Alice's measurements in the standard basis.
When {Alice's outcome $\left\vert 0\right\rangle _{0}$ occurs} in the probability $N b_0^2$,
the teleportation can be accomplished by two local  von Neumann measurement on quNits $1$ and $2$.
One the other hand, Alice can recover the initial state by a joint unitary operation,  $B_{21}=\sum_{ j=0}^{ N-1} U_{1}^{(0,j\oplus1) \dag} \otimes | j \rangle _{2}\langle  j |$, on her two systems, when the task of teleportation fails in the  probability $1-N b_0^2$.
\\

 \section{Summary}\label{summ}

We present a scheme for nondestructive probabilistic teleportation of high-dimensional quantum states.
A partial entangled pure state severs as the quantum channel, whose smallest coefficient  determines the successful probability of  exactly teleporting a state.
With the aid of an auxiliary particle, Alice can  recover her initial state to be teleported when teleportation fails.
Compared to the existing results \cite{JOSAB2020a,JOSAB2020b}, our protocol only requires the sender, Alice, to have the ability to perform bipartite operations, while the dimension of the ancilla needs to be the same as the state to be teleported.
In addition, the ancilla acts as a quantum apparatus to measure Alice's system.
The process of resuming the initial state can be regarded as erasing information recorded in the ancilla.

As the {following} research, it is a fundamental problem to explore the roles of quantum correlations in our four-party procedure,
which is a fundamental problem in quantum information.
In addition, the relation between  our protocol and the theory of extracting information from a quantum system by multiple observers \cite{bergou2013extracting} would be interesting,
since the quantum correlations in the latter are studied in many works \cite{pang2013sequential,ZhangJHQINP,ZhangJHPRA}.
While we focus here on the teleportation using quantum channels  with  the same dimension as the state to be teleported,
it is a natural extension to apply  the present ideas in the case with different dimension \cite{Zhang2021}.
And finally, we hope  {that} the process can be implemented in laboratories with the help of the techniques recently developed in optical systems \cite{2019Quantum,2020Experimental}.

%




\end{document}